\documentclass[pra,twocolumn, showpacs,a4paper,superscriptaddress]{revtex4}

\usepackage{latexsym}
\usepackage{amsmath}
\usepackage{amssymb}
\usepackage{amsfonts}
\usepackage{comment}
\usepackage{graphicx}
\usepackage{bbm}
\usepackage{color}

\newcommand{\ld}{\,\textrm{log$_2$}\,}
\newcommand{\tr}{\,\textrm{tr}\,}
\newcommand{\im}{\textrm{i} }
\newcommand{\psucc}{p_\textrm{succ}}
\newcommand{\perr}{p_\textrm{err}}
\newcommand{\iabt}{I_{A\&B}^{\textrm{total}}}
\newcommand{\iaet}{I_{A\&E}^{\textrm{total}}}

\begin{document}

\title{The Singapore Protocol: Incoherent Eavesdropping Attacks}

\author{Janet Anders\footnote{email: janet@qipc.org}}
\affiliation{Quantum Information Technology Lab, Department of Physics, 
National University of Singapore, Singapore 117542}

\author{Hui Khoon Ng\footnote{Present address: Department of Physics,
    California Institute of Technology, Pasadena, CA 91125, USA.}} 
\affiliation{Quantum Information Technology Lab, Department of Physics,
  National University of Singapore, Singapore 117542} 
\affiliation{Applied Physics Lab, DSO National Laboratories, Singapore 118230}

\author{Berthold-Georg Englert}
\affiliation{Quantum Information Technology Lab, Department of Physics,  
National University of Singapore, Singapore 117542}

\author{Shiang Yong Looi }
\affiliation{Quantum Information Technology Lab, Department of Physics, 
National University of Singapore, Singapore 117542}

\date{\today}

\begin{abstract}
We thoroughly analyse the novel quantum key distribution protocol introduced
recently in quant-ph/0412075, which is based on minimal qubit tomography. We
examine the efficiency of the protocol for a whole range of noise parameters
and present a general analysis of incoherent eavesdropping attacks with
arbitrarily many steps in the iterative key generation process. The comparison
with the tomographic 6-state protocol shows that our protocol has a higher
efficiency (up to 20\%) and ensures the security of the established key even
for noise parameters far beyond the 6-state protocol's noise threshold. 

\end{abstract}

\pacs{}

\maketitle

\subsection{Introduction}

With the growing interest in quantum information theory, quantum cryptography
has become a key research area.  One way of ensuring secure communication is
to establish a secret key between the two communication partners, \emph{Alice}
and \emph{Bob}, with which they can later encode and  decode secret
messages. The distribution of the key can be done using quantum mechanical
systems where the laws of physics guarantee the security, in marked contrast
to classical schemes which rely on the complexity of mathematical
problems. Quantum key distribution could thus replace conventional public key
cryptosystems, which can be broken in polynomial time by quantum algorithms as
soon as suitable quantum computers are available.  

Most quantum key distribution schemes discussed at the current stage of
research are based on the BB84 \cite{BB84} or the E91 \cite{E91}
protocol. These schemes operate in only a subspace  of the whole qubit state
space, and so they allow the eavesdropper, \emph{Eve}, unnecessary freedom to
make use of the undetected portion of the Hilbert space. This power is denied
to her in fully tomographic key distribution protocols, such as those of
Ref.\,\cite{TomoCrypt}, and in particular by the ``minimal qubit tomography 
protocol'' of Ref.\,\cite{TetraCrypt} that has become known colloquially as the
\emph{Singapore protocol}. It relies heavily on the minimal qubit tomography
(MQT) discussed in Ref.\,\cite{MQT}

In Ref.\,\cite{TetraCrypt} it was shown that the maximum theoretical
efficiency of a quantum key distribution protocol using the MQT measurement is
maximally $\ld \frac{4}{3} = 0.415$; this is less than the efficiencies of the
non-tomographic BB84 and  E91 protocol which are $\frac{1}{2}$ and
$\frac{2}{9}$, respectively. However, the difference between the mutual
information between Alice and Bob (A\&B) and the mutual information between
Eve and either Alice or Bob (CK-yield) promises to be higher than in the
non-tomographic protocols \cite{BB84-disc}.  Moreover, the MQT measurement has
the potential for a significantly higher key yield than the comparable
tomographic 6-state protocol introduced in Ref.\,\cite{B} and discussed in
Refs.\,\cite{B-PG,GW}, which has an efficiency of only $\frac{1}{3}$.   

Using the MQT measurement, one possible way to generate a secret key from the
correlated measurement outcomes was proposed in Ref.\,\cite{TetraCrypt}. The
resulting quantum key distribution protocol recovers 0.4 key-bits per qubit
pair, or 96.4\% of the potential efficiency in the noise-free
case. It is the objective of the present paper to give a detailed account of
how the analysis is carried out that yields the thresholds stated in
Ref.\,\cite{TetraCrypt}. A quantum key-distribution scheme that uses POVMs
with tetrahedron structure is also investigated by Renes in
Ref.\,\cite{Renes}. This protocol differs from the Singapore protocol by a
less efficient key generation procedure that does not fully exploit the
potential of the MQT measurement. 

In Sec.\,\ref{sec:MQT} we give a brief overview of the MQT measurement and
review the key generation in Sec.\,\ref{sec:Singapore}. In
Sec.\,\ref{sec:constraints} we will discuss the constraints on Eve's
eavesdropping imposed by the tomographic nature of the MQT measurement.  We
then investigate, in Sec.\,\ref{sec:eavesdropping}, the incoherent attacks
available to Eve when she exploits the classical information transmitted
between the communication partners during the key generation. Finally, in
Sec.\,\ref{sec:security} we examine the security of the protocol  against such
attacks and obtain the noise threshold stated in Ref.\,\cite{TetraCrypt}.

\subsection{Minimal qubit tomography} \label{sec:MQT}

We suppose Alice and Bob want to establish a secret key and they use a
provider that distributes entangled qubit pairs for private communication. As
advertised, each will receive one qubit of the pair. Since real communication
channels do not usually preserve the signal perfectly, Alice and Bob have to
deal with the fact that they will receive a distorted state. Let  Alice and
Bob agree to accept only a mixed state $\rho_{A\&B}$ consisting of the ideal,
perfectly anti-correlated singlet $ |s \rangle\langle s| = {1 \over 4}
\left(\mathbbm{1} - \vec   \sigma_{A} \cdot \vec \sigma_{B} \right),$ and
white, unbiased noise, weighted  with a noise parameter  $\epsilon$. The
two-qubit state that A\&B will share is thus  
\begin{eqnarray}\label{rhoAB}
  \rho_{A\&B}(\epsilon) =(1-\epsilon)|s\rangle\langle s| 
  +\frac{\epsilon}{4}\mathbbm{1},
\end{eqnarray}
where $\epsilon$  ranges from  $\epsilon =0$ (no noise) to $\epsilon= 1$
(nothing but noise). In practical situations, the class of acceptable sources
should in fact be  chosen in accordance with  the experimental setup which
produces the singlet and the properties of the transmission line used (fibre,
air, ...). But for the sake of simplicity, we impose the above standard
criterion which was also used in the tomographic protocols of
Ref.\,\cite{TomoCrypt}.   

The tomographically complete 6-state protocol was analyzed for the above
scenario in Ref.\,\cite{AccInf}. In this protocol, Alice and Bob each
performs a measurement of a randomly chosen Pauli operator, $\sigma_x,
\sigma_y$ or $ \sigma_z$, resulting in six outcome probabilities. However,
this is not a \emph{minimal} tomography  since  only four outcome
probabilities are needed to specify the state of a qubit completely. The
optimal qubit tomography POVM with the minimum number of four elements is of
tetrahedron geometry as shown in Ref.\,\cite{MQT}; that is the POVM operators
can be written in the form      
\begin{equation}\label{eq:POVM}
  P_k = \frac{1}{4} \left( \mathbbm{1} + \vec{t}_k \cdot \vec{\sigma}\right)
  \quad\mbox{for}\ k =1, 2, 3, 4,
\end{equation}
where the vectors $\vec t_k$ point to the corners of a tetrahedron inscribed
in the Bloch sphere; see Fig.\,1 in Ref.\,\cite{MQT} for an illustration. The
four vectors are linearly dependent,  
\begin{equation}
        \sum_{k=1}^4\vec{t}_k=0,
\end{equation}
 with the scalar product
\begin{equation}
        \vec{t}_k\cdot\vec{t}_l=\frac{4}{3}\delta_{kl}-\frac{1}{3}\quad
        \mbox{for}\ k,l =1, 2, 3, 4,
\end{equation}
and fulfill the dyadic completeness relation
\begin{equation}
        \frac{3}{4}\sum_{k=1}^4 ~  \vec{t}_k\,\vec{t}_k=\tensor{1}.
\end{equation}  

Let Alice and Bob each measure the tetrahedron POVM of Eq.\,\eqref{eq:POVM} on
many copies of a two-qubit state $\rho$. The resulting joint probabilities of
the measurement are then given by $p_{kl} = \tr[\rho ~P_k \, Q_l],$ with $P_k$
denoting Alice's POVM elements and $Q_l$ Bob's, chosen so that their
tetrahedrons are perfectly aligned (if they chose a non-zero angle
between their tetrahedrons they would lose the perfect anti-correlations
introduced by the singlet). To verify that they indeed received the state
$\rho_{A\&B}$ of Eq.\,\eqref{rhoAB}, Alice and Bob sacrifice a fraction of
their data and announce them publicly to determine the joint probabilities
$p_{kl}$ of Alice measuring $k$ and Bob $l$. They check their results for
statistical independence and are able to reconstruct the original state by 
\begin{eqnarray}\label{2qbstate}
\rho=\sum_{k,l=1}^4 ~(6 P_k -\mathbbm{1})~ p_{kl}~ (6 Q_l-\mathbbm{1}).
\end{eqnarray} 
Naturally, after a finite number of measurements, Alice and Bob cannot infer
the values of the $p_{kl}$ exactly, but they can estimate them rather
reliably. A discussion of the quality of such estimates was given in
Ref.\,\cite{MQT} for the single qubit case where it  was also shown that the
measurement of a randomly chosen qubit state  with the tetrahedron POVM will
on average lead to the best (\emph{optimal}) estimate of the state's Pauli
vector. Finally A\&B compare whether the predicted state of
Eq.\,\eqref{2qbstate} is consistent with Eq.\,\eqref{rhoAB}. They will only
use the provider if this is the case.   

Given  the shared state is $\rho_{A\&B}(\epsilon)$ for some $\epsilon$, their
joint probabilities $p_{kl}$ will be    
\begin{equation}
  \label{pkl} 
    p_{kl}=\frac{4-\epsilon}{48}(1-\delta_{kl})
    +\frac{\epsilon}{16}\delta_{kl} \quad \mbox{for } k,l =1, 2, 3, 4,
\end{equation}
and the accessible information that A\&B can establish between each other is
given by 
\begin{equation}
  \label{eq:potentialmutinfo}
  I^{\textrm{access}}_{A\&B}(\epsilon) =
  \left(1-\frac{\epsilon}{4}\right)\ld \frac{4-\epsilon}{3} 
  + \frac{\epsilon}{4}\ld \epsilon,   
\end{equation}
where we used the definition of the classical mutual information for a
probability distribution $\{ p_{kl}\}_{k,l}$  
\begin{equation}
I = \sum_{kl}~ p_{kl} ~ \ld \frac{p_{kl}}{\sum_{k'} ~ p_{k'l} 
~  \sum_{l'}  ~p_{kl'}}. 
\end{equation}
Note that in the noise-free case the accessible information,
$I^{\textrm{access}}_{A\&B} (0) = 0.415$, is substantially  higher (by 24.5\%)
than the corresponding value of $\frac{1}{3}$ for the tomographically complete
6-state protocol.

\subsection{The Singapore protocol}\label{sec:Singapore}

From now on, we will refer to the possible measurement outcomes as  A, B, C, D
for $k = 1,2,3,4$, respectively, symbolizing a click in the $k$-th POVM
detector. To generate a key from their correlated sequences, Alice and Bob
have to communicate classically. We use the two-way key generation scheme
proposed in Ref.\,\cite{TetraCrypt} which leads to a mutual information of
$I_{A\&B}(0) = 0.4$, sufficiently close to the maximally accessible
information of  $I^{\textrm{access}}_{A\&B}(0) =  0.415$. The scheme has
a simple structure and can be easily implemented on a computer.  We give a
brief description of the key generation scheme followed by a more detailed
analysis considering the presence of noise which will be relevant for the
eavesdropping discussion. We refer the reader to the original paper
\cite{TetraCrypt} for more details on the key generation scheme.   

Let us first consider the noise-free case ($\epsilon=0$). Alice publicly
announces two randomly chosen positions of her sequence for which she has the
same letter. With probability $\frac{2}{3}$, Bob has different letters at
these positions. He then groups the possible outcomes A, B, C, D in two
groups, one containing the two letters he received and one with the remaining
two letters. He randomly assigns the values 0 and 1 to the two groups and
announces these groupings. Bob does not reveal which group his letters belong
to, but  since A\&B's measurement outcomes are perfectly anti-correlated, they
can both write down the value of the group which contains Alice's letter and
thus generate a key-bit.  With probability $\frac{1}{3}$  Bob has the same
letter in the two positions Alice announced. In this case, he states this fact
and A\&B each write their corresponding letter in a new sequence. \emph{The
  above procedure is repeated iteratively with the new sequences thus
  generated.}   

In the presence of noise, the key sequence generated with the above scheme
will contain errors with a rate dependent on $\epsilon$. For the original
letter  sequence (first iteration), the probability of Alice and Bob receiving
the same letter is then 
non-zero, see Eq.\,\eqref{pkl},  
\begin{equation} \label{eq:ps}
  p_s\,(\epsilon) = \frac{\epsilon}{4},
\end{equation}
and Bob receives one of the other three letters with probability 
\begin{equation} \label{eq:pd}
  p_d\,(\epsilon) = \frac{4-\epsilon}{12}.
\end{equation}
With \emph{a priori} probability $\frac{1}{4}$ Alice announces the positions
of two occurrences of the letter A. Then, Bob's corresponding two letters
occur with the probabilities given in Table \ref{tab:Bob2letters},
where M.P. denotes the marginal probabilities.  
\begin{table}[t]
  \begin{center}
    \begin{tabular}[c]{cccccc|c}
      \hline
      \hline
      \multicolumn{2}{c}{} & \multicolumn{4}{c}{Bob's 2nd letter}&\\
      \multicolumn{2}{c}{} & A & B & C & D & M.P.\\
      \hline
      & A 
      & $\frac{1}{4} p_s^2$  & $\frac{1}{4}p_s p_d$ 
      & $\frac{1}{4}p_s p_d$ & $\frac{1}{4}p_s p_d$ 
      & $\frac{\epsilon}{16}$\\
      Bob's 
      & B
      & $\frac{1}{4}p_s p_d$ & $\frac{1}{4}p_d^2$ 
      & $\frac{1}{4}p_d^2$   & $\frac{1}{4}p_d^2$ 
      & $\frac{4-\epsilon}{48}$\\
      1st letter
      & C
      & $\frac{1}{4}p_s p_d$ & $\frac{1}{4}p_d^2$ 
      & $\frac{1}{4}p_d^2$   & $\frac{1}{4}p_d^2$ 
      & $\frac{4-\epsilon}{48}$\\
      & D
      & $\frac{1}{4}p_s p_d$ & $\frac{1}{4}p_d^2$ 
      & $\frac{1}{4}p_d^2$   & $\frac{1}{4}p_d^2$ 
      & $\frac{4-\epsilon}{48}$\\[1ex]
      \cline{2-7}
      & M.P.
      & $\frac{\epsilon}{16}$ & $\frac{4-\epsilon}{48}$
      & $\frac{4-\epsilon}{48}$ & $\frac{4-\epsilon}{48}$
      & $\frac{1}{4}$\\[1ex]
      \hline
      \hline
    \end{tabular}
    \caption{\label{tab:Bob2letters}Bob's two letters given that Alice
      announced two positions where she got the outcome A.}  
  \end{center}
\end{table}
The conversion into one key-bit will occur when Bob's letters are unequal,
i.e.  in all off-diagonal cases. Similarly, we can construct the probability
tables for Alice's other choices of letters. Let us denote the probability of
successfully generating one key-bit from one letter pair by $\psucc$. For the
first iteration it is then  
\begin{equation} \label{eq:psucc}
  \psucc^{(1)}\,(\epsilon) = 6 p_d \left( p_s +p_d \right) =
  \frac{\left(4-\epsilon\right)\left(2+\epsilon\right)}{12}.  
\end{equation}
However,  these successfully generated key-bits will contain errors. The
probability that a generated key-bit is wrong is given by  
\begin{equation} 
  \perr^{(1)}~(\epsilon)= \frac{6 p_s p_d}{\psucc^{(1)} } 
  =\frac{p_s}{p_s +  p_d}=\frac{3\epsilon}{4+2\epsilon}, 
\end{equation}
accounting for the equally likely cases that Bob writes $0$ and Alice $1$ and
vice versa. The mutual information of the key itself is thus less than unity
and depends on $\perr^{(1)}$, which is nonzero for nonzero $\epsilon,$  
\begin{eqnarray}
  I _{\textrm{key}}\left(\perr^{(1)}\right) 
  &=& 1 + \perr^{(1)} ~\ld \perr^{(1)} \\
  && +  (1-\perr^{(1)})~\ld \left[1-\perr^{(1)}\right]. \nonumber   
\end{eqnarray} 
Let us regard the mutual information as a cryptographic resource that Alice
and Bob can use later to extract a perfectly correlated key. We are therefore
interested in the \emph{expectation value of the mutual information} which
A\&B share \emph{per qubit pair}. This expectation value is the product of the
mutual information of the generated key-bit and the probability that this
key-bit was actually generated ($\psucc^{(1)}$), divided by the number of
qubit pairs needed (2 in the first iteration) to obtain the key-bit.
\begin{eqnarray} \label{eq:IAB}
  I_{A\&B}^{(1)}(\epsilon) 
  &=& \frac{\psucc^{(1)}}{2} \, I_{\textrm{key}}\left(\perr^{(1)}\right)\\
  &=& \frac{\left(4-\epsilon\right)}{48} \left((4-\epsilon)\ld
  \frac{4-\epsilon}{2+\epsilon} +  3\epsilon\ld \frac{3\epsilon}{2+\epsilon}
  \nonumber \right).  
\end{eqnarray}

To deduce similar results for further iterations we first study the properties
of the recycled sequences. The second iteration can again be characterized by
two probabilities $p_s'$ and $p_d'$ defined with analogous meanings as $p_s$
and $p_d$ for the original sequence. The probability $p_s'$ is given by the
probability of Bob receiving the same letter as Alice in the original sequence
twice $\left( {p_{s}}^2\right)$, divided by the total probability of keeping
letters in the first iteration, i.e. failure in generating a key-bit,
$\left(1-\psucc^{(1)} \right)$,   
\begin{eqnarray}
  p_s'=\frac{\left(\frac{\epsilon}{4}\right)^2}{
  \left(\frac{\epsilon}{4}\right)^2+3\left(\frac{4-\epsilon}{12}\right)^2}.
\end{eqnarray}
Similarly, $p_d'$ is given by
\begin{eqnarray}
  p_d'=\frac{\left(\frac{4-\epsilon}{12}\right)^2}{
    \left(\frac{\epsilon}{4}\right)^2+3\left(\frac{4-\epsilon}{12}\right)^2} 
  =\frac{1-p_s'}{3}. 
\end{eqnarray}
Upon defining $\epsilon'$ in accordance with $p_s'= \frac{\epsilon'}{4}$ and
$p_d'=\frac{4-\epsilon'}{12}$, we can carry the analysis for the first
iteration over to the next iteration by replacing $\epsilon$ by
$\epsilon'$. The relation between $\epsilon'$ and $\epsilon$ is most compactly
stated in the form   
\begin{eqnarray}\label{eq:eprime}
  \frac{3 \epsilon'}{4-\epsilon'} = \left(\frac{3 \epsilon}{4-\epsilon}
  \right)^2,   
\end{eqnarray}
showing that the noise reduces quadratically with each iteration step. All
further iterations can thus be analyzed in the same way, each time replacing
the noise  $\epsilon$ of the previous iteration by the new noise parameter $
\epsilon'$.  

In particular, the probability of successfully generating a key-bit in the
$n$-th iteration is, similarly to Eq.\,\eqref{eq:psucc}, given by
$q^{(n)}=\left(4-\epsilon^{(n)}\right)\left(2+\epsilon^{(n)}\right)/12$, where
$\epsilon^{(n)}$ denotes the noise parameter in the $n$-th iteration. The
conditional probability  $\psucc^{(n)}$ of a key-bit being generated in the
$n$-th iteration, after failure in the previous $n-1$ iterations is then given
by 
\begin{eqnarray}\label{eq:psuccn}
  \psucc^{(n)} = q^{(n)}  \prod_{m=1}^{n-1} \left(1-q^{(m)} \right).
\end{eqnarray}
The error probability per key-bit $\perr^{(n)}$ for the $n$-th iteration can
easily be written as
\begin{eqnarray}
  \perr^{(n)} = \frac{3\epsilon^{(n)}}{4+2\epsilon^{(n)}}  
  = \left[1+\left(\frac{4-\epsilon}{3\epsilon}
    \right)^{2^{n-1}}\right]^{-1},
\end{eqnarray}
which uses Eq.\,\eqref{eq:eprime}. The contribution to $\iabt $ in the $n$-th
iteration is given by $I_{A\&B}^{(n)}(\epsilon)= 2^{-n} \, \psucc^{(n)} \,
I_{\textrm{key}}\left(\perr^{(n)}\right)$, and the overall expectation value
of the mutual information per qubit pair in the limit of infinitely many
iterations is thus   
\begin{eqnarray} \label{eq:IABtotal}
  \iabt (\epsilon)
  &=&\sum_{n=1}^\infty \,\frac{\psucc^{(n)}}{2^n} \,
  I_{\textrm{key}}\left(\perr^{(n)}\right). 
\end{eqnarray}
This quantity serves as our figure of merit for the comparison with the
6-state protocol. One should however keep in mind that it is an average over
the various key-bit sequences of the successive iterations. Each of them has
different noise properties which must be taken into account when the data are
processed further by a privacy amplification procedure.

\begin{figure}[t] 
\begin{center}  
{\includegraphics[width=0.45\textwidth]{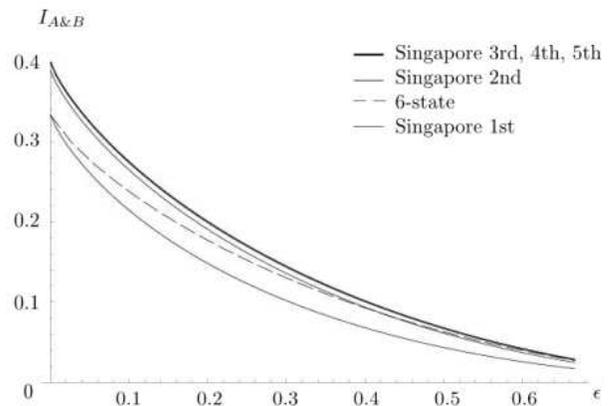}}
\caption{\label{fig:IAB} 
Total mutual information between Alice and Bob for the Singapore protocol for
the 1st to 5th iteration in comparison with the tomographic 6-state
protocol. The plotted average mutual information of Alice and Bob in the
6-state protocol represents the function
$I^{\textrm{6-state}}_{A\&B}(\epsilon) = \frac{1}{6} \left( \epsilon \ld
  \epsilon  + (2-\epsilon) \ld [2-\epsilon] \right).$ Note, that the total
mutual information up to  the 3rd, 4th and 5th iteration are so close that
they overlap and the 3rd iteration is already a very good approximation of $
\iabt $. The plot covers the range $0< \epsilon < \frac{2}{3}$, i.e., the
$\epsilon$-values for which  $\rho_{A\&B}$ of Eq.\,\eqref{rhoAB} is not
separable. }
\end{center}
\end{figure}

In the noiseless case ($\epsilon =0$), the Singapore protocol yields a mutual
information $I_{A\&B}^{(1)}(0)= {1\over 3}$ for the first iteration. This is
as much as one can get in the 6-state protocol; but here we can improve the
efficiency by continuing the key generation with the left-over sequences,
e.g. $I_{A\&B}^{(1)}(0) + I_{A\&B}^{(2)}(0)=  {1 \over 3} + {1 \over 18 } =
0.389$ up to the second iteration and so on, with the limiting value of
0.4. When Alice and Bob share the complete mixture ($\epsilon =1$) we find the
expected result of $\iabt (1) \propto \ld [1] = 0.$ The numerical plots of the
total mutual information for Alice and Bob when terminating the key generation
after one, two, ..., five iterations are shown in Fig.\,\ref{fig:IAB}. The
comparison with the 6-state protocol shows that the mutual information
obtained in the Singapore protocol is larger from the third iteration
onwards. Alice and Bob do not need to perform more than three key-generation
iterations as the third iteration already comes so close to the limiting value
that the benefit of further iterations will be less then 0.01\%. The total
gain compared to the 6-state protocol is $\Delta I_{A\&B} = 0.066$ or 20\% in
the noiseless case and vanishes for $\epsilon \ge \frac{2}{3}$, when
$\rho_{A\&B}$ of Eq.\,\eqref{rhoAB} is separable. This larger efficiency is
expected since the Singapore protocol is minimally tomographic and does not
waste as much information as the 6-state protocol.   

Our first conclusion is, therefore, that the Singapore protocol is
considerably more efficient than its obvious competitor, the tomographic
6-state protocol, and offers an alternative way for establishing a secret key
between the communication partners. We will now discuss possible incoherent
eavesdropping attacks on the Singapore protocol and find the noise thresholds
below which the security of the protocol is guaranteed.   

\subsection{Constraints on Eve's eavesdropping}\label{sec:constraints}

In Sec.\,\ref{sec:MQT}  Alice and Bob received a two-qubit state sent by a
provider.  We must however assume that this provider is not trustworthy and
eager to know A\&B's secret. We hence identify the provider as the most
dangerous eavesdropper (Eve) possible, and give her full control over the
source. In the \emph{worst case scenario}, Eve is smarter with her technology
and can even replace the usually noisy channel between Alice and Bob by a
perfect one. She is then in the position to entangle an additional ancilla to
each qubit pair she sends, and the disturbances she causes by doing so would
imitate noise. But Alice and Bob perform a complete tomography of the shared
state and since they agree to use the channel only when the noise has the
properties reflected in Eq.\,\eqref{rhoAB}, they can greatly restrict how
Eve can entangle ancillas, and later on deduce the shared key.    

Eve will prepare a 3-party \emph{pure} state because she has no advantage from
creating a mixed state and thus introducing classical noise herself. We
decompose the state as  
\begin{equation}
  \label{Sepsilon}
  | S_{\epsilon} \rangle = \sum_{l=1}^4 |l \bar l \rangle | E_l \rangle,
\end{equation}
where the (unnormalized) states $| E_l \rangle $ represent Eve's ancilla. The
pure states $|l\bar l \rangle $ belong to Alice and Bob and form a
non-orthogonal basis $\{|l \bar l \rangle\}_{l=1}^4$, where the tetrahedron
states $| l \rangle $ and $| \bar l \rangle $ are defined as 
\begin{eqnarray}\label{eq:tetra}
  \begin{aligned}
    |l \rangle \langle l | 
    &= \frac{1}{2} \left(\mathbbm{1} + \vec t_l \cdot \vec \sigma \right), \\
    |\bar l \rangle \langle \bar l | 
    &= \frac{1}{2} \left(\mathbbm{1} - \vec t_l \cdot \vec \sigma \right),  \\
  \end{aligned} \quad l = 1, 2, 3, 4,
\end{eqnarray}
with the phase-conventions 
\begin{eqnarray}
  \langle l|k\rangle = \langle \bar k|\bar l\rangle \quad \mbox{and} \quad 
  \langle l|\bar k\rangle = -  \langle k|\bar l\rangle,
\end{eqnarray}
the second of which implied by the first. Some useful relations of these
states can be found in the Appendix. Note, that the decomposition of
Eq.\,\eqref{Sepsilon} allows for exactly four components on Eve's side,
together forming Eve's ancilla state. Each ancilla can thus be represented by
a maximally four-dimensional system so that the ancilla can be regarded as
another qubit pair.  

Eve's ancilla state is however restricted by the condition that A\&B must
receive the two-qubit state $\rho_{A\&B}$ of Eq.\,\eqref{rhoAB} which they
check by comparing their outcomes of the tomographic measurement. Assuming
that all the noise originates in Eve's eavesdropping attempt, the 3-party
state $|S_{\epsilon} \rangle$ must be such that   
\begin{equation} 
\tr_\textrm{Eve}\left[|S_\epsilon\rangle\langle S_\epsilon|\right]
=\rho_{A\&B}(\epsilon).
\end{equation}
Let us rewrite $\rho_{A\&B}$  using the expansion of the identity matrix in
Eq.\,\eqref{eq:identity} and the decomposition of the singlet in
Eq.\,\eqref{eq:e}, to get  
\begin{eqnarray}
  \rho_{A\&B}(\epsilon)
  &=& \frac{1}{8}\sum_{l,k=1}^4 |l \bar l \rangle
  \langle k \bar k |  \left(1 - \frac{3}{2} \epsilon  
+ 3 \epsilon \delta_{lk}\right). 
\end{eqnarray}
This expression immediately gives the following constraints on Eve's ancilla
state  
\begin{eqnarray}\label{anccon}
    \langle E_k | E_l \rangle = \frac{2-3\epsilon}{16}
    +\frac{3\epsilon}{8}\delta_{kl} ~ \mbox{ for } ~ k, l = 1, ...,4.
\end{eqnarray}
For $\epsilon = 0$, the  $|E_l \rangle$ are identical and Eve cannot extract
any information. For $\epsilon =\frac{2}{3}$, the scalar products in
Eq.\,\eqref{anccon} show the orthogonality of the  $|E_l\rangle$ which implies
that $\rho_{A\&B}$ is separable. Indeed, Alice and Bob share a separable
Werner state   
\begin{eqnarray} 
    \rho_{A\&B}\left(\frac{2}{3}\right) =\frac{1}{4}\sum_{l=1}^4~|l\bar
    l\rangle\langle l\bar l| 
    =\frac{1}{4}\left(1-\frac{1}{3} \vec \sigma_A \cdot \vec \sigma_B\right), 
\end{eqnarray}
and thus all correlations are classical. 

In Ref.\,\cite{pyramids} it was shown that, given the constraints of
Eq.\,\eqref{anccon}, the most general state Eve can construct, up to unitary
equivalence, can be written as    
\begin{eqnarray}\label{Sepsilon1} 
  |S_{\epsilon} \rangle &=& 
  \alpha | s_{12} \rangle | s_{34} \rangle +  \beta |s_{13} \rangle |s_{24}
  \rangle, 
\end{eqnarray}
where the first qubit is held by Alice, the second by Bob and the third and
fourth by Eve. The amplitudes $\alpha$ and $\beta$ must now be chosen so that
Eve's ancilla satisfies Eq.\,\eqref{anccon}. By using Eq.\,\eqref{Sepsilon}
and Eq.\,\eqref{Sepsilon1} we find Eve's  states to be  
\begin{eqnarray}
  | E_k \rangle &=& \alpha \frac{1}{2\sqrt{2}}  | s \rangle 
  -\frac{\beta}{2}\left(|\bar k k\rangle +\frac{1}{2}|k \bar  k \rangle\right),
\end{eqnarray}
and evaluating the scalar products $\langle E_k|E_l \rangle$ and comparing with
Eq.\,\eqref{anccon}  we deduce the constraints on the parameters $\alpha$ and
$\beta$, 
\begin{eqnarray}\label{eq:ab1} 
    |\beta|^2 = \epsilon \qquad\mbox{and}\qquad \left|\alpha + 
    \frac{\beta}{2}\right|^2 = 1 -\frac{3}{4}\epsilon.
\end{eqnarray}
Since we have a freedom of global phase for $|S_\epsilon\rangle$ we can choose
$\beta$ to be real, i.e. $\beta=\sqrt\epsilon$. The only free parameter  is
then the phase $\phi$ in 
\begin{equation}\label{eq:phi}
\alpha + \frac{\beta}{2} = e^{\im\phi}\sqrt{1-\frac{3\epsilon}{4}}.
\end{equation}    
 
Eve wants to guess Alice's key-bit and constructs a state $\rho^{(k)}_E$ for
each outcome $k$ Alice could measure regardless of Bob's result. These
conditional ancilla states are  
\begin{eqnarray}\label{eq:rhok}
  \rho^{(k)}_E &=& \tr_{A\&B} \left[P_k ~ |S_\epsilon\rangle\langle
    S_\epsilon|  \right],\\ 
  &=&\frac{|\beta |^2}{8} |\bar k\bar k \rangle\langle \bar k\bar k|
  \nonumber\\ 
  && + \, \frac{1}{4} \left(\alpha |s\rangle - \frac{\beta}{\sqrt 2} |\bar k
    k\rangle \right)   
  \left(\alpha^* \langle s| - \frac{\beta^*}{\sqrt 2} \langle \bar k k|
    \right),\nonumber 
\end{eqnarray}
where $P_k$ is the POVM element for Alice measuring $k$. Note, that all
$\rho_E^k$ are subnormalized to ${1 \over 4}$, the \emph{a priori} probability
that Alice will measure a particular $k$. 

Owing to the symmetry between Alice and Bob, the ancilla states conditional to
Bob's measurement results are unitarily equivalent to these
$\rho^{(k)}_E$. Therefore, it does not matter whether Eve tries to learn
Alice's measurement results or Bob's.

\subsection{Incoherent Eavesdropping attacks}\label{sec:eavesdropping}

Let us summarize what we have found so far. For each qubit pair that Eve sends
to A\&B she will keep a qubit pair (the ancilla) for herself. In the first
iteration of the key generation scheme Alice and Bob will use the measurement 
outcomes of two qubit pairs. Eve has thus two corresponding ancillas which she
can measure to guess A\&B's generated key-bit. For the second iteration Eve
will have four, for the third she will have eight ancillas and so on.   

We suppose Eve has no means of storing her ancillas until classical
communication between Alice and Bob is done. She therefore has to measure them
individually as she creates them, without being able to include the classical
information in her measurement.  Her measurement can be optimized such as to
maximize her mutual information with Alice. We will call her optimal strategy
for doing this an \emph{incoherent attack} as opposed to a coherent (joint)
measurement performed on the bunch of ancillas  correlated through the key
generation process. The optimal  POVM for the $\rho_E^{(k)}$ of
Eq.\,\eqref{eq:rhok} was found using the iterative procedure in
Ref.\,\cite{AccInf} and is analogous to the optimal POVM for the 6-state
protocol given there. The 4-member POVM consists of the projectors  
\begin{eqnarray} \label{eq:el}
  M_l&=&|e_l\rangle\langle e_l|,  \nonumber\\  
  |e_l\rangle&=&{1 + \sqrt 3 e^{- \im \phi} \over 2} |s \rangle
  + \sqrt{\frac{ 3}{2}} e^{- \im \phi} | \bar l l \rangle,
\end{eqnarray}
for $l=1, 2, 3, 4$, where $\phi$ is the phase of Eq.\,\eqref{eq:phi}, and the
$M_l$ obey $\sum_{l=1}^4 \, M_l = \mathbbm{1}$. Note that the POVM is
independent of the noise parameter, $\epsilon$. 

Interestingly, in the interval $0 < \epsilon < \bar{\epsilon}$ where
$\bar{\epsilon} = 0.1725$ (obtained numerically by solving a transcendental
equation), it was found that a 5-member POVM gives a slightly larger mutual
information then the 4-member POVM (less than 1\% larger).  The fifth element
has the following expression 
\begin{eqnarray} \label{eq:e5}
  M_5&=&|e_5\rangle\langle e_5| \;\; \mbox{where} \;\; |e_5\rangle = \sqrt{2
    \mu - 4 \mu^2} \sum_{l=1}^4 |e_l\rangle\ 
\end{eqnarray}
where $\mu$ is a function of $\epsilon$ and $0 \leq \mu \leq 1/2$. All the
other $\{|e_l\rangle\}_{l=1}^4$ have to be modified in the following manner
\begin{eqnarray} \label{eq:sum_povm5}
  |e_j\rangle \to \left( |e_j\rangle  - \mu \sum_{l=1}^4 |e_l\rangle\ \right) 
\end{eqnarray}
to ensure the elements sum up to identity. For the purpose of finding the
noise threshold, we can just use the simpler 4-member POVM since as we shall
see later, the noise threshold is always much larger than $\bar{\epsilon}$ for
any iteration.  

With the 4-member POVM, the joint probabilities $q_{kl}$ of Alice measuring
$k$ and Eve measuring $l$ are given by the same expression as Alice and Bob's
joint probabilities in Eq.\,\eqref{pkl} with $\epsilon$ replaced by a new
noise parameter $\eta$,quantifying the noise between Alice and Eve, with  
\begin{equation}\label{eq:eta}
  \eta(\epsilon)=\left(\sqrt{1-\frac{3\epsilon}{4}}
    -\sqrt\frac{3\epsilon}{4}\right)^2. 
\end{equation}
Note that for $\epsilon=0$ the noise between Alice and Eve reaches a maximum
($\eta=1$) and when $\epsilon=\frac{2}{3}$ there is no noise between Alice and
Eve ($\eta=0$). 

We define the probabilities $q_s$ ($q_d$) for Alice and Eve having the same (a
particular different) measurement result in analogy to the $p_s$ ($p_d$) in
Eq.\,\eqref{eq:ps} (Eq.\,\eqref{eq:pd}) by   
\begin{eqnarray} \label{eq:qsqd}
         q_s(\eta) = q_{kk}=\frac{\eta}{4} 
        \quad \textrm{and} \quad q_d(\eta)= q_{k\neq l}=\frac{4-\eta}{12}.
\end{eqnarray}
When a key-bit is generated between Alice and Bob the joint probabilities
between Alice's and Eve's results are as given in Table \ref{tab:AE} where we
assumed that Bob grouped  $\framebox{AB} = 0$ and $\framebox{CD} =1$
(similarly for other groupings). Note, that the probabilities in Table
\ref{tab:AE} are again more anti-correlated than they are correlated since
$q_s \le q_d$ for all $\eta$. We compute the mutual information between Alice
and Eve from this table of probabilities. The middle column does not
contribute and the mutual information becomes   
\begin{eqnarray}\label{eq:IAE} \nonumber
  I_{A\&E}^{(1)}(\epsilon)&=&\frac{\psucc^{(1)}}{2} \left\{
    (q_s^2+q_d^2) \ld\left[\frac{q_s^2+q_d^2}{2}\right]\right.  \\ 
  && + \left.
    4 q_d^2  \ld\left[q_d^2\right] + 2 q_s q_d \ld\left[q_s q_d\right] 
  \right. \nonumber \\ && - \left.
    (q_s^2 + 3 q_d^2) \ld\left[\frac{q_s^2 + 3 q_d^2}{4}\right]
  \right.  \\ && - \left.
    2 q_d (q_s + q_d) \ld\left[\frac{q_d(q_s + q_d)}{2}\right]
  \right\}. \nonumber
\end{eqnarray} 
where the prefactor is the same as in Eq.\,\eqref{eq:IAB}. This value gives
the upper bound to the amount of information Eve can obtain about the key
generated in the first iteration. It is valid for incoherent attacks only, by
whatever suitable method Eve might employ to extract Alice's key-bit. 

\begin{table}[t]
  \begin{tabular}{ccccccc|c}
    \hline
    \hline
    \multicolumn{2}{c}{}
    & \multicolumn{5}{c}{Eve's combinations} & \\[1ex]
    \multicolumn{2}{c}{}
    & AA    & AB    & AC, AD        & CD    & CC    & \\
    \multicolumn{2}{c}{}
    & BB    & BA    & CA, DA        &DC     & DD    & \\
    \multicolumn{2}{c}{}
    &       &       & BC, BD        &       &       & \\
    \multicolumn{2}{c}{}
    &       &       & CB, DB        &       &       & M.P.\\
    \hline
    Alice's 
    & 0    
    &$\frac{q_s^2+q_d^2}{4}$        
    &$\frac{q_sq_d}{2}$ &$\frac{q_d(q_s+q_d)}{4}$ 
    &$\frac{q_d^2}{2}$                      
    &$\frac{q_d^2}{2}$              
    & 1/2\\[1ex]
    key-bit 
    &1 
    &$\frac{q_d^2}{2}$                      
    &$\frac{q_d^2}{2}$  &$\frac{q_d(q_d+q_s)}{4}$  
    &$\frac{q_sq_d}{2}$  
    &$\frac{q_d^2+q_s^2}{4}$  
    & 1/2 \\[1ex]
    \cline{2-8} 
    &M.P. 
    &$X$ 
    &$Y$
    &$Y$
    &$Y$
    &$X$
    & 1\\ 
    \hline
    \hline
  \end{tabular}
  \caption{\label{tab:AE}Joint probabilities between Alice and Eve with the
    marginals $X = \left(q_s^2+ 3 q_d^2\right)/4$ and $Y= q_d(q_d+q_s)/2$, for
    the grouping AB = 0 and CD =1.} 
\end{table}

In the $n$-th iteration Eve will have $2^n$ qubit pairs available for
measurement. She measures all qubit pairs individually  and gets a sequence of
$2^n$ letters with A, B, C and D occurring   $n_A, n_B, n_C$ and $n_D = 2^n
-n_A-n_B-n_C $ times, respectively. In all announced  positions which
contribute in the key generation of a key-bit, Alice always has the same
letter, say A. The probability of Eve measuring a sequence which contains
$n_A$ times A is given by, with $q_s$ and $q_d$ from Eq.\,\eqref{eq:qsqd},  
\begin{equation} 
        q_n^{\textrm{A}}(n_A) = \frac{1}{4} \,q_s^{n_A} \,q_d^{2^n -n_A}.
\end{equation}
If Bob grouped $\framebox{AB} = 0$ and $\framebox{CD} =1$, say, the
probability of Eve getting a particular distribution $\{n_A, n_B, n_C, n_D\}$
and Alice having the key-bit  0 is then   
\begin{eqnarray} \nonumber
  q_n^0 (n_A, n_B,) 
  &=&  q_n^{\textrm{A}} (n_A) + q_n^{\textrm{B}} (n_B),\\
  &=&\frac{q_d^{2^n }}{4}  \left[  \left( \frac{q_s}{q_d}\right)^{n_A} 
    + \left(\frac{q_s}{q_d}\right)^{n_B}\right]. 
\end{eqnarray}
and similarly for group 1. The marginal probabilities for Eve getting a
distribution $\{n_A, n_B, n_C, n_D\}$ is then  
\begin{eqnarray} \nonumber
  q_n(n_A, n_B,n_C,n_D) 
  &=&  q_n^{0} (n_A, n_B) + q_n^1 (n_C,n_D),\\   
  &=&\frac{q_d^{2^n }}{4} \sum_{J=A}^D \, \left(\frac{q_s}{q_d}\right)^{n_J}.
\end{eqnarray}
The number of sequences with a particular distribution $\{n_A, n_B, n_C,
n_D\}$ is  
\begin{equation}
  \binom{2^n}{n_A,n_B,n_C,n_D} 
  = \frac{2^n! ~ \delta_{2^n, n_A+n_B+n_C+n_D}}{n_A!~n_B!~ n_C!~ n_D!},
\end{equation}
and Alice's marginals $q_n^k$ for  $k=0,1$ turn out correctly  
\begin{eqnarray}\nonumber
  q_n^k &=& \sum_{n_A, n_B, n_C, n_D} \binom{2^n}{n_A,n_B,n_C,n_D} ~ q_n^k
  (n_A, n_B,n_C,n_D) \\          
  &=& \frac{1}{2}.  
\end{eqnarray}
The contribution to the  mutual information that Eve shares with Alice from
the $n$-th iteration can now be calculated,
\begin{eqnarray}\label{eq:IAEn} \nonumber 
  I_{A\&E}^{(n)}(\epsilon)&=&\frac{\psucc^{(n)}}{2^n} ~
  \sum_{k=0}^1 ~~ \sum_{n_A,n_B,n_C,n_D=0}^{2^n} \\ \nonumber 
  && \times \binom{2^n}{n_A,n_B,n_C,n_D}  \, q_n^k (n_A, n_B,n_C,n_D) \, \\
  && \times \ld \left[ \frac{q_n^k (n_A, n_B,n_C,n_D)}{q_n (n_A, n_B,n_C,n_D)
  \, q_n^k }\right], 
\end{eqnarray}
where $\psucc^{(n)}$ is again the probability of success in the $n$-th
iteration (Eq.\,\eqref{eq:psuccn}) and the factor of $2^{-n}$ gives the
mutual information per qubit pair used. The total mutual information that Eve
can reach if Alice and Bob perform infinitely many iterations is then 
\begin{eqnarray}\label{eq:IAEtotal}
  \iaet (\epsilon) &=& \sum_{n=1}^{\infty} \,I_{A\&E}^{(n)} (\epsilon).
\end{eqnarray}
In the noise-free case the mutual information of Eve vanishes $\iaet
(0)\propto  \ld \left[ 1\right] =0$. This is clear since the channel between
Alice and Bob is perfect and the channel between Alice and Eve is completely 
noisy ($\eta=1$).

\subsection{Security} \label{sec:security}

According to the Csisz\'ar-K\"orner Theorem in Ref.\,\cite{CK}, Alice and Bob
are able to share a secret key provided their mutual information $\iabt $
exceeds the mutual informations shared between Eve and one of the
communication partners $\iaet$ and $I_{B\&E}^{\textrm{total}}$.  The CK-yield
is then  
\begin{equation}\label{eq:CKyield} 
Y_{CK} = \iabt -\iaet. 
\end{equation}
The CK-yield determines the length of the secure key Alice and Bob can obtain
from the generated raw key of length $L$; namely the length of the secure key
will maximally be $Y_{CK} \, L$ (for one-way communication). The intersection
between $\iabt$ and $\iaet$ thus gives the final noise threshold below which
a secret key between Alice and Bob can be generated by one-way communication
that relies on error correction codes.       

\begin{figure}[t]
  \begin{center}
    {\includegraphics[width=0.4\textwidth]{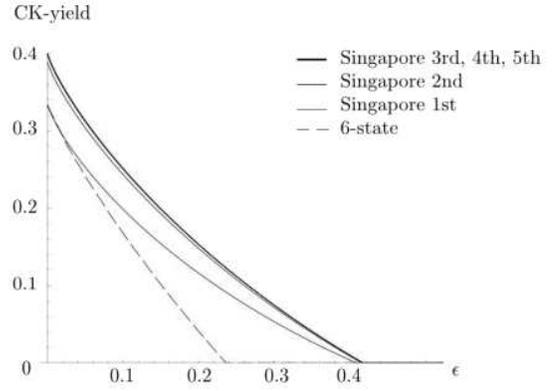}}
    \caption{\label{fig:yield}Yield for the Singapore protocol for the 1st to
      5th iteration in comparison with the tomographic 6-state protocol. For
      the 6-state protocol we used the mutual information between Alice and
      Eve obtained in Ref.\,\cite{AccInf}. The yields for the 3rd, 4th and 5th
      iteration overlap too, as they did in Fig.\,\ref{fig:IAB} and the latter
      can be regarded as a very  close approximation to the yield in the limit
      of infinitely many iterations. }   
\end{center}
 \end{figure}

In Fig.\,\ref{fig:yield} we plot the CK-yield for the 6-state protocol and
different iterations of the Singapore protocol. Observe that the yield of the
Singapore protocol is distinctly larger than the yield of the tomographic
6-state protocol from the second iteration onwards. For $\epsilon = 0$ the
gain is already 20\%  and it increases significantly for larger noise
parameters. Further, the noise threshold for the first iteration of the
Singapore protocol is at $\epsilon = 0.409$ and increases to $\epsilon =
0.417$  for the 3rd iteration. Additional iterations will raise this value in
the 4th decimal place, so that $\epsilon = 0.417$ can count as the maximum
noise Alice and Bob can accept when establishing a secure key with the
Singapore  protocol. In contrast, the 6-state protocol has its noise threshold
at the much smaller value of $\epsilon = 0.236.$ Compared to the 6-state
protocol the noise threshold of the Singapore protocol is remarkable 76.7\%
higher. This result is our key observation in this paper. Given a number of
qubit pairs, measuring the tetrahedron POVM and using the above key-generation
scheme thus leads to a raw key substantially longer than the one that could be
produced by the 6-state protocol. Additionally, Alice and Bob are still able
to share a secret key when the noise level exceeds the 6-state threshold by a
lot.

\subsection{Discussion and conclusions}

The analysis of the eavesdropping attacks was carried out for the original
protocols, without any error-correction or privacy amplification schemes. It
is thus a  comparison between the Singapore protocol and the 6-state protocol
in their pure form only up to the extraction of a \emph{raw key}. We have seen
the inherent potential of both protocols and analysed in detail how much
mutual information the communication partners Alice and Bob can establish
between each other when using the Singapore protocol. It turns out that A\&B
can already  stop the key generation after the third iteration without losing
much, since the mutual information up to the third iteration is already very
close to the limiting value of infinitely many iterations. The comparison with
the 6-state protocol showed that the efficiency of the Singapore protocol is
up to 20\% larger than in the 6-state protocol. 

We continued our discussion by constructing incoherent eavesdropping attacks
under the following assumptions: 
1) A source controlled by the eavesdropper Eve distributes the singlet state,
mixed with unbiased, white noise scaled by the noise parameter $\epsilon$.
2) Eve is the cause for all noise; and 
3) Eve can not store her ancillas and is thus not able to incorporate knowledge
of the classical communication between Alice and Bob when measuring her
ancillas. Additionally she is constrained to perform only individual
attacks and cannot measure correlated ancillas in a joint
measurement. Condition 1) is equivalent to the scenario where Alice sends a
qubit in a state orthogonal to one of the tetrahedron states from
Eq.\,\eqref{eq:tetra}, each with probability $\frac{1}{4}$. Eve could then
intercept the traveling qubit and produce an optimal clone and an
anti-clone. Here she also keeps two qubits which she can measure and Bob
receives a disturbed state. Together with condition 2) this leads to the
\emph{worst case scenario} for Alice and Bob, where Eve can realize optimal
cloning. This is reasonable since we are interested in absolute security
statements which rely only on the laws of physics and not on technical
abilities of Eve. On the other hand, we assumed condition 3) which is clearly
a relaxation to this strictness. But this is still a valid constraint given
that modern technology has not developed reliable quantum storage systems and
it is not yet feasible to perform joint measurements on demand. However, we
have also analysed coherent eavesdropping attacks on the Singapore protocol as
discussed in Ref.\,\cite{TetraCrypt} and will present a detailed and extended
report in due time.      

Our discussion did assume throughout that Alice and Bob share a state of the
form given in Eq.\,\eqref{rhoAB}. A natural and open question is then how well
will the Singapore protocol perform if the state differs from the above,
e.g. if the noise is somehow biased, and how does it then compare to the
6-state protocol? Another issue worth addressing is the possible use that Eve
can make of the information gained when Alice and Bob perform a privacy
amplification or key purification by other means. However, under the
conditions 1) - 3), the Singapore protocol provides an efficient alternative
for generating a secret key between two communication partners. The measurement
of the tetrahedron POVM is from a practical point of view as feasible as the
comparable tomographic  6-state measurement (see Ref.\,\cite{OneLoop}) and the
efficiency and security under incoherent attacks is significantly higher.   

\acknowledgements

We gratefully acknowledge inspiring discussions with S. M. Assad and
W. K. Chua. H. K. Ng would like to thank the Defence Science and Technology
Agency (DSTA) of Singapore for their financial support. J. Anders gratefully
acknowledges the financial and personal support of the Gottlieb Daimler und
Karl Benz-Stiftung. This work was supported by A$^*$Star Grant
No. 012-104-0040 and by NUS Grant WBS: R-144-000-109-112.

\appendix

\subsection*{Appendix}

\setcounter{equation}{0}
\renewcommand{\theequation}{A\arabic{equation}}

It is expedient to re-express the tetrahedron states of Eq.\,\eqref{eq:tetra}
in terms of the  POVM elements $P_l$ of Eq.\,\eqref{eq:POVM} by 
\begin{eqnarray}\label{eq:tetra2}
  \begin{aligned}
    |l \rangle \langle l | &= 2~ P_l, \\
    |\bar l \rangle \langle \bar l | &= \mathbbm{1} - 2~ P_l.
  \end{aligned} \quad l = 1, 2, 3, 4,
\end{eqnarray}  
Then the scalar products of the tetrahedron states are given by 
\begin{eqnarray}
  \begin{aligned}
    |\langle l|k\rangle|^2 =4\tr[P_l\,P_k]
    &=\frac{1}{3}(1+2\delta_{kl}),\\   
    |\langle l|\bar k\rangle|^2 =2\tr[P_l~ (\mathbbm{1} -2 P_k)]
    &=\frac{1}{3}(2 - 2\delta_{kl}).
  \end{aligned}
\end{eqnarray}
The singlet $|s \rangle $ can be written in terms of the $|l \rangle$ and $|
\bar l \rangle$ as   
\begin{eqnarray} \label{eq:e}
  |s \rangle &=& \frac{1}{\sqrt 2} (|l \bar l \rangle - |\bar l l \rangle)
  \,\, \mbox{for \emph{any} } l = 1, 2, 3, 4,
\end{eqnarray}
or
\begin{eqnarray}
  |s \rangle &=& \frac{1}{\sqrt 8} \sum_{l=1}^4 ~ |l \bar l \rangle 
  = -\frac{1}{\sqrt 8} \sum_{l=1}^4 ~ |\bar l l \rangle.
\end{eqnarray} 
Furthermore, we expand the identity in terms of $| l \bar l \rangle$ and
the singlet,  
\begin{equation}\label{eq:identity}
  \mathbbm{1} = \frac{3}{2}\sum_{l=1}^4  |l \bar l \rangle \langle l\bar l |
  - 2 | s \rangle \langle s|. 
\end{equation}


\begin{thebibliography}{99}
\bibitem{BB84-disc}
For a discussion of the security of the BB84-protocol under incoherent
attacks, see for example Refs.\,\cite{GW,FGGNP}. These references discuss the
security under different assumptions about the setup of the source and how the
noise is introduced. The results obtained are not directly comparable with the
results for the Singapore protocol. Protocols with partial tomography, such as
the BB84 or E91, can be analyzed in a similar manner (work in progress).
\end{thebibliography}
\end{document}